\documentclass[aps, prd, twocolumn, amsmath, floats,floatfix, 10pt, superscriptaddress, nofootinbib,
showkeys]{revtex4}

\DeclareRobustCommand{\VAN}[3]{#2}
\let\VANthebibliography\thebibliography
\def\thebibliography{\DeclareRobustCommand{\VAN}[3]{##3}\VANthebibliography}

\usepackage{amssymb}
\usepackage{amsmath}
\usepackage{verbatim}
\usepackage{mathrsfs}
\usepackage{amsfonts}
\usepackage{latexsym}
\usepackage{epsfig}
\usepackage{color}
\usepackage{graphicx,subfigure}
\usepackage{units}
\usepackage{envmath}
\usepackage{natbib}
\usepackage{ctable}
\usepackage[T1]{fontenc}
\usepackage[colorlinks=true, linkcolor=red, citecolor=blue, urlcolor=black]{hyperref}
\begin{document}

%%%%%%%%%%%%%%%%%%
%%%   MACROS   %%%
%%%%%%%%%%%%%%%%%%

\definecolor{orange}{rgb}{0.9,0.45,0}

\def\CovDev{D}
\def\Res{{\mathcal R}}
\def\Gammaflat{\hat \Gamma}
\def\metricflat{\hat \gamma}
\def\Dflat{\hat {\mathcal D}}
\def\part_n{\partial_\perp}

%=== Definition for abbreviations ===
\def\Lie{\mathcal{L}}
\def\A{\mathcal{X}}
\def\Aphi{\A_{\phi}}
\def\hAphi{\hat{\A}_{\phi}}
\def\E{\mathcal{E}}
\def\Ham{\mathcal{H}}
\def\M{\mathcal{M}}
\def\R{\mathcal{R}}
\def\p{\partial}

\def\hg{\hat{\gamma}}
\def\hA{\hat{A}}
\def\hD{\hat{D}}
\def\hE{\hat{E}}
\def\hR{\hat{R}}
\def\hcA{\hat{\mathcal{A}}}
\def\hDelt{\hat{\triangle}}

\def\na{\nabla}
\def\dif{{\rm{d}}}
\def\non{\nonumber}
\newcommand{\erf}{\textrm{erf}}
%====================================

\renewcommand{\t}{\times}

\long\def\symbolfootnote[#1]#2{\begingroup%
\def\thefootnote{\fnsymbol{footnote}}\footnote[#1]{#2}\endgroup}

%%%%%%%%%%%%%%%%%
%%%   TITLE   %%%
%%%%%%%%%%%%%%%%%

\title{On the effect of angular momentum on the prompt cusp formation via the gravitational collapse}
 
\author{Antonino Del Popolo}
\email{antonino.delpopolo@unict.it}
\affiliation{Dipartimento di Fisica e Astronomia, University of Catania, Viale Andrea Doria 6, 95125 Catania, Italy}
\affiliation{Institute of Astronomy, Russian Academy of Sciences, Pyatnitskaya str. 48, 119017 Moscow, Russia}
  
\author{Saeed Fakhry}
\email{s\_fakhry@sbu.ac.ir}
\affiliation{Department of Physics, Shahid Beheshti University, G. C., Evin, Tehran 19839, Iran}
\affiliation{PDAT Laboratory, Department of Physics, K.N. Toosi University of Technology, P.O. Box 15875-4416, Tehran, Iran}

%%%%%%%%%%%%%%%%
%%%   DATE   %%%
%%%%%%%%%%%%%%%%

\today

%%%%%%%%%%%%%%%%%%%%
%%%   ABSTRACT   %%%
%%%%%%%%%%%%%%%%%%%%

\begin{abstract} 
In this work, we extend the model proposed by White in \cite{2022MNRAS.517L..46W} concerning the post-collapse evolution of density peaks while considering the role of angular momentum. On a time scale smaller than the peak collapse, $t_0$, the inner regions of the peak reach the equilibrium forming a cuspy profile, as in White's paper, but the power-law density profile is flatter, namely $\rho \propto r^{-1.52}$, using the specific angular momentum $J$ obtained in theoretical models of how it evolves in CDM universes, namely $J \propto M^{2/3}$. The previous result shows how angular momentum influences the slope of the density profile, and how a slightly flatter profile obtained in high-resolution numerical simulations, namely $\rho \propto r^{\alpha}$ ($\alpha \simeq -1.5$) can be reobtained. Similarly to simulations, in our model adiabatic contraction was not taken into account. This means that more comprehensive simulations could give different values for the slope of the density profile, similar to an improvement of our model.
\end{abstract}

%%%%%%%%%%%%%%%%
%%%   PACS   %%%
%%%%%%%%%%%%%%%%

\keywords{
Dark Matter -- Gravitational Collapse -- Galactic Halos
}

%%%%%%%%%%%%%%%%%%%%%%
%%%   MAKE TITLE   %%%
%%%%%%%%%%%%%%%%%%%%%%

\maketitle

\vspace{0.8cm}

%%%%%%%%%%%%%%%%%%%%%%
\section{Introduction}
%%%%%%%%%%%%%%%%%%%%%%

Dark matter halos are nonlinear hierarchical structures whose formation and evolution are predicted in the cosmological perturbation theory \citep{Hiotelis2013}. The initial stages of the formation of these structures can be attributed to the physical conditions under which the primordial density fluctuations can be separated from the expansion of the Universe and collapse due to the self-gravitational force. Dark matter halos are a suitable and fundamental framework for studying nonlinear gravitational collapse in the Universe. Therefore, the post-collapse evolutionary stages of dark matter halos play an essential role in explaining their local properties, see, e.g. \citep[][]{DelPopolo2000,2009ApJ...698.2093D, 2010AdAst2010E...5D,DelPopolo2014,2017MNRAS.470.2410R, DelPopolo2021}.

Accordingly, in recent years, many high-resolution simulations of collapse and post-collapse evolution of dark matter halos have been performed, see, e.g. \citep[][]{2010ApJ...723L.195I, 2013JCAP...04..009A, 2014ApJ...788...27I, 2017MNRAS.471.4687A, 2018MNRAS.473.4339O, 2018PhRvD..97d1303D, 00001254, 2023MNRAS.518.3509D}. The outcomes of these simulations demonstrate that shortly after the collapse of the initial density peaks, the central regions of dark matter halos can be well described by the power-law density profile $\rho(r)=Ar^{\gamma}$ with $\gamma \approx -1.5$. 
In this relation, $A$ is constant, and its value is estimated for each dark matter halo from the characteristic scale and collapse time of the relevant density peak \citep{2023MNRAS.518.3509D}. 
We want to recall that the high-resolution simulation by \citep{2023MNRAS.518.3509D} are not taking into account baryons, and this implies that several important physical effects, like, for example, adiabatic contraction are not taken into account. This implies that the result $\gamma \approx -1.5$ could be modified by those effects.
On the other hand, the dynamics of hierarchical structures in dark matter models, except for those with self-interactions, indicate that the galactic halos in the earliest stages of their post-collapse evolution start to grow instantaneously in their size and mass and ultimately reach a uniform non-power-law density distribution \citep{1996ApJ...462..563N, 2004MNRAS.349.1039N, 2009A&A...502..733D, 2011JCAP...07..014D}.

Dark matter-only simulations of galaxy-type and cluster-sized halos indicate that the effective slope of halo density profiles at the smallest resolution radii must be shallower than $\gamma \approx -1.5$, see, e.g. \citep[][]{2008MNRAS.391.1685S, 2012MNRAS.425.2169G, 2014JCAP...07..019D}. However, the slope of the density profile may return to a steep state due to the resistance of the initial cusp in the central regions of dark matter halos, see, e.g. \citep[][]{2023MNRAS.518.3509D}. Although many numerical studies have been conducted in the post-collapse evolution of initial density peaks, the black-box nature of simulations does not explain the formation of prompt cusps with a power-law index $\gamma \approx -1.5$.

In order to provide a logical description of the shape of the formed halos, the first theoretical model was presented in \cite{1972ApJ...176....1G}, in which an initial density peak prone to collapse is considered as a perturbed point-like mass in a dust-filled collisionless Einstein-de Sitter Universe. The results of this study show that dark matter halos with a power-law index $\gamma = -9/4$ are created when the surrounding matter falls into the perturbation. After that, in \cite{1984ApJ...281....1F}, a more general approach was proposed by considering spherical collapse in purely radial orbits, which did not predict the power-law index describing dark matter halos as $\gamma>-2$. However, assuming purely radial orbits to describe the complex collapse conditions seemed simplistic. Accordingly, it was shown in \cite{1996clss.conf.....S} that the consideration of randomly oriented orbits with non-zero angular momentum leads to providing an interval for the power-law index as $0>\gamma>-3$. The mentioned studies all agree that the orbital period of the circle in the radius encompassing mass $M$ is proportional to the time of the fall of the halo shell that surrounded mass $M$ in the early Universe. Also, in \cite{2013MNRAS.432.1103L}, a more complete analytical model was presented in explaining the density profile of dark matter halos, which describes the relationship between the fall time and the halo structure.

Despite their valuable results, the aforementioned studies cannot provide information on the instantaneous formation of the cusp during the earliest stages of post-collapse evolution of the initial density peaks, because the process of instantaneous formation of cusps requires a different description of the fall of the shells into the halo in a suitable timescale. In this regard, \cite{2022MNRAS.517L..46W} has presented an analytical model for the post-collapse evolution of initial density peaks in a collisionless dust-filled Universe. The results of this study exhibit that on instantaneous time scales compared to the collapse time of the initial density peaks, the innermost regions of the formed halos are consistent with the density profile of adiabatic cusps with a power distribution index $\gamma=-12/7$. The power-law index value obtained by \cite{2022MNRAS.517L..46W} is not compatible with the relatively flatter corresponding value of $\gamma \approx -1.5$ obtained from high-resolution numerical simulations. Notably, in the analysis presented in \cite{2022MNRAS.517L..46W}, the effect of angular momentum is not included, which can significantly reduce the difference between analytical approaches and high-resolution simulations.

In this work, we focus on studying the effect of angular momentum on the prompt cusp formed during the post-collapse evolution of initial density peaks. In this respect, the outline of the work is
as follows. In Sec.\,\ref{sec:ii}, We discuss a theoretical framework for the gravitational collapse from the initial density peaks and evaluate its post-collapse evolutionary stages in the presence of angular momentum. Also, in Sec.\,\ref{sec:iii}, we discuss the results obtained in this work and compare them with those extracted from the previous studies. Finally, in Sec.\,\ref{sec:iv}, we summarize our findings.
%%%%%%%%%%%%%%%%%%%%%%%%%%%%%%%%%%%%%%%%%%%%%%%%
\section{Theoretical model of gravitational collapse}\label{sec:ii}
As mentioned in the previous section, dark matter halos are nonlinear structures formed by the gravitational collapse of peaks in overdensity regions in a dust-filled collisionless Universe. In the earliest stages of the formation of initial density peaks, cosmological perturbation theory estimates the local density of the neighborhood of the peaks as
\begin{equation}
    \rho(q, t)=\Bar{\rho}(t)[1+\delta(q, t)]=\Bar{\rho}(t)\left[1+\delta_{\rm sc}(t)\left(1-\dfrac{q^{2}}{6R^{2}}\right)\right],
\end{equation}
where $q=r/a$ is a Lagrangian radial coordinate, $\Bar{\rho}=1/(6\pi G t^{2})=\Bar{\rho_{0}}(t/t_{0})^{-2}=\Bar{\rho_{0}}a^{-3}$ is the mean density of the background, $a=(t/t_{0})^{2/3}$ is the cosmological expansion factor, $R=\sqrt{|\delta/\bigtriangledown^{2} \delta|}$ is a characteristic Lagrangian scale for the peak, and $\delta_{\rm sc}=1.686 a$ is the critical value of linear overdensities. Also, the index $``0"$ represents the evaluated values of the quantities at the collapse time of the central region of the peak to infinite density. The mass enclosed by a radius of $q$ is determined as $M=4\pi\Bar{\rho}_{0}q^{3}/3=M_{0}(q/R)^{3}$, where $M_{0}= 4\pi\Bar{\rho}_{0}R^{3}/3$. Hence, the average overdensity within radius $q$ can be specified by the following relation
\begin{eqnarray}
    \Bar{\delta}(q, t)=1.686 a \left(1-\dfrac{q^{2}}{10R^{2}}\right)\hspace{1.15cm}\nonumber\\ 
    =1.686 a \left[1-0.1\left(\dfrac{M}{M_{0}}\right)^{2/3}\right].
\end{eqnarray}

\begin{figure*}
\includegraphics[width=6.5cm]{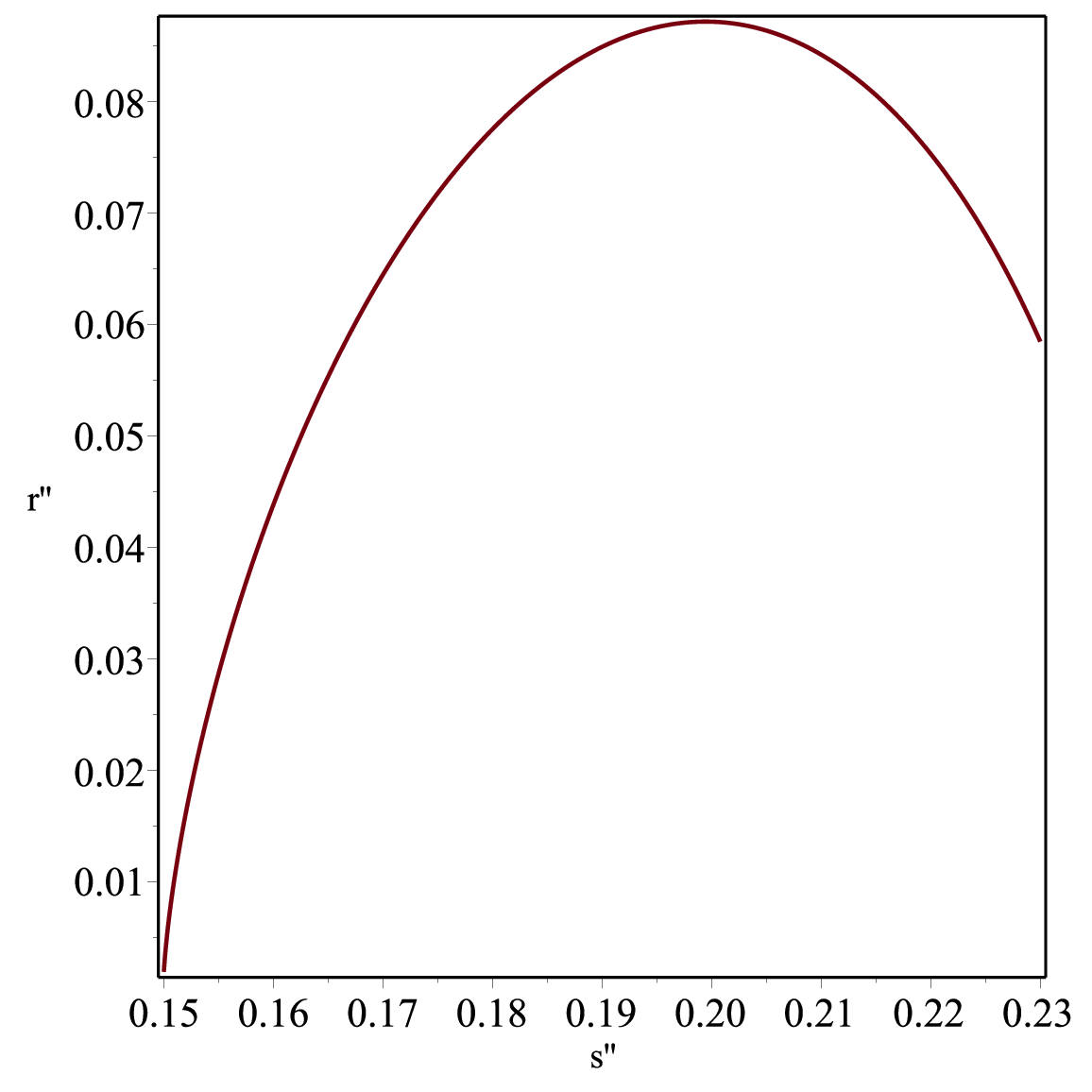}
\includegraphics[width=6.5cm]{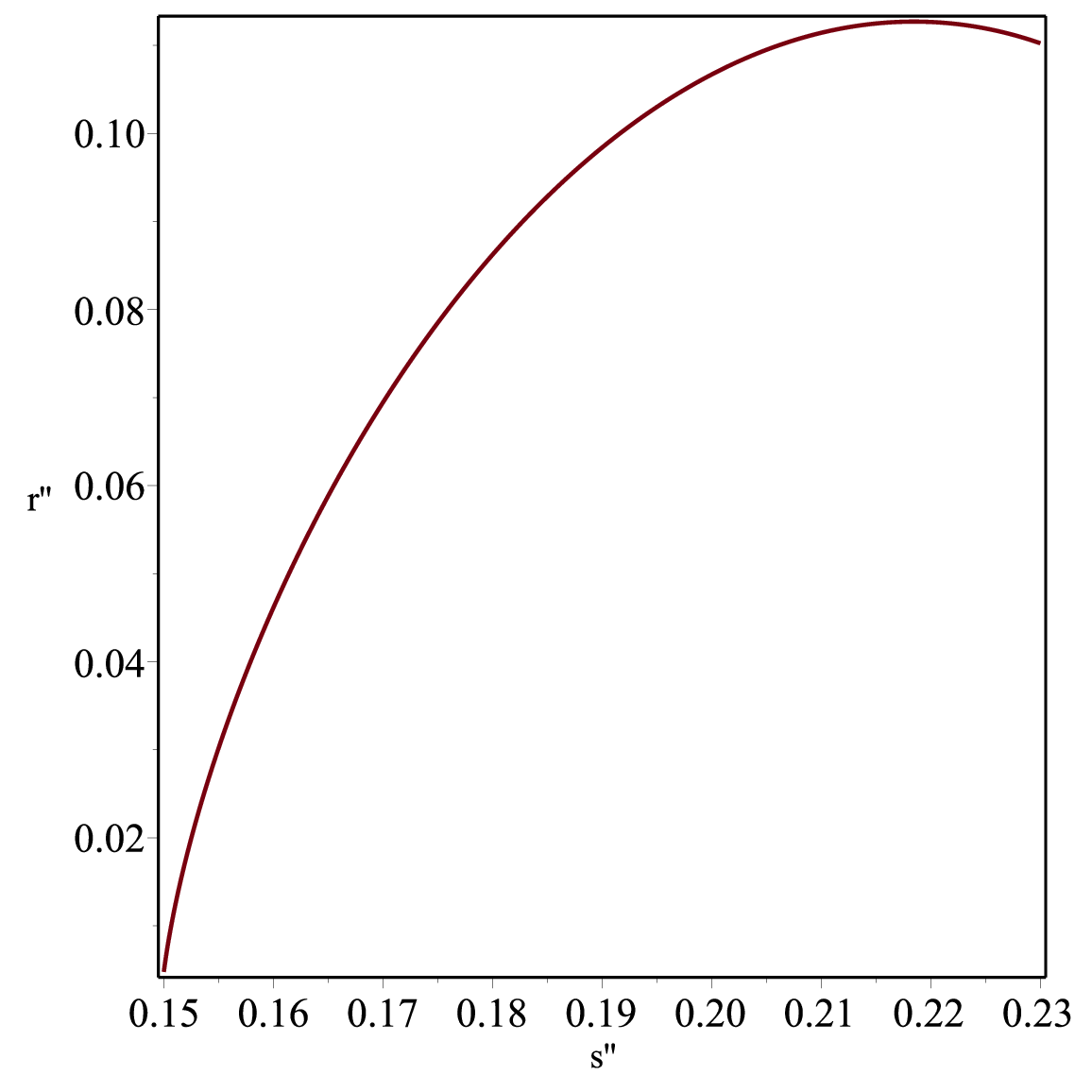}
\caption{Post-collapse trajectory. Left panel: solution without angular momentum, $J=0$. Right panel: solution taking into account angular momentum, and $m'=0.5$}
\label{fig:example_figure}
\end{figure*}

Due to maintaining the radial hierarchy in mass shells, their collapse happens just when the mass enclosed in the shell reaches the critical value of overdensities $1.686$. Therefore, the shell collapse time for the lowest order in $M/M_{0}$ can be calculated as
\begin{equation}\label{collapse_time}
    t_{\rm c}(M)=t_{0}\left[1+0.15\left(\dfrac{M}{M_{0}}\right)^{2/3}\right].
\end{equation}
At times earlier than $t_{\rm c}(M)$, considering the non-zero angular momentum, the infall velocity of the shell is
\begin{equation}\label{vel_rel}
    \dfrac{1}{2}\left(\dfrac{dr}{dt}\right)^{2}=
    \dfrac{GM}{r} + \int\dfrac{L^{2}}{M^{2}r^{3}}dr,
\end{equation}
where $G$ is the gravitational constant and $L$ is the angular momentum. 

As discussed in several papers, e.g. \cite{2009ApJ...698.2093D}, two sources of angular momentum are involved in structure formation: (1) derived from bulk streaming motions and (2) produced by random tangential motions. The former, the ordered angular momentum, arises due to tidal torques experienced by protohalos \citep{Peebles1969,White1984,Catelan1996}. The latter, the random angular
momentum \cite{Ryden1987}, is connected to random velocities (non-radial motions).  

Several studies have concluded that larger amounts of angular momentum, ordered or random, lead to shallower inner density profiles, see, e.g. \citep[][]{Ryden1987,AvilaReese1998,Nusser2001,Hiotelis2002,LeDelliou2003,Ascasibar2004,Williams2004,2009ApJ...698.2093D,DelPopolo2012,DelPopolo2014a,Polisensky2015}. In our case, in these smooth peaks, there are no random motions. 
%
%For the sake of simplicity, we solely consider ordered angular momentum in this study. However, it should be noted that %including the effect of random angular momentum leads to an increase in the flattening of the density profile }
%\citep{DelPopolo2000,DelPopolo2006,2009ApJ...698.2093D

The previous relation describes how the angular momentum of typical particles in a shell scales with Lagrangian radius under the assumption of a uniform tidal field across the considered region.
As mentioned above, the ordered angular momentum in the CDM model has been studied by several authors \citep{Peebles1969,White1984,Catelan1996}. The specific angular momentum of each mass element is defined as $J=L/M$, and takes the form $J=J_{0}(M/M_{\rm J})^{2/3}$ \cite{Catelan1996,Romanowsky2012,Lagos2016}, where $J_{0}$ and $M_{\rm J}$ are characteristic angular momentum and mass.

 Hence, solving Eq.\,\eqref{vel_rel} yields the following relation
\begin{eqnarray}\label{solution1}
    r^{3/2}F(r,M)=3(GM_{\rm J})^{2}\left(\dfrac{M}{M_{\rm J}}\right)^{1/2}(t_{\rm c}-t),
\end{eqnarray}
where 
\begin{eqnarray}
    F(r,M)=\sqrt{2GM_{\rm J}-\dfrac{J_{0}^{2}}{r}\left(\dfrac{M}{M_{\rm J}}\right)^{1/3}}\times\hspace{1.7cm}\nonumber\\
    \left[GM_{\rm J}+\dfrac{J_{0}^{2}}{r}\left(\dfrac{M}{M_{\rm J}}\right)^{1/3}\right].
\end{eqnarray}
Equivalently, one can write Eq.\,\eqref{solution1} as
\begin{eqnarray} \label{solutioni}
     r^{3/2}F(r,M)=3(GM_{\rm J})^{2}\left(\dfrac{M}{M_{\rm J}}\right)^{1/2}t_{0}\times\hspace{1.7cm}\nonumber\\
     \left[0.15\left(\dfrac{M}{M_{0}}\right)^{2/3}-\dfrac{\Delta t}{t_{0}}\right],
\end{eqnarray}
where $\Delta t=t-t_{0}$. Note that Eq.\,\eqref{solutioni} has been obtained from Eq.\,\eqref{solution1}, substituting the value of $t_c$ given by Eq.\,\eqref{collapse_time}. Here, we can assume that Eq.\,\eqref{collapse_time} is still valid because, as we will see, angular momentum slightly changes the spherical collapse whose collapse time is given by Eq. \eqref{collapse_time}.

After simplifying this expression, it takes the following form,
\begin{eqnarray}\label{r_over_R}
     \left(\dfrac{r}{R}\right)^{3/2}F(r,M)
    =\sqrt{2}(GM_{\rm J})^{3/2}\left(\dfrac{M}{M_{0}}\right)^{1/2}\times \hspace{1.5cm}\nonumber \\
    \left[0.15\left(\dfrac{M}{M_{0}}\right)^{2/3}-\dfrac{\Delta t}{t_{0}}\right],
\end{eqnarray}
in which $0<\Delta t/t_{0}<0.15(M/M_{0})^{2/3}$ and $M<M_{0}$ are valid for a short time interval after the initial collapse of the peak. A collapsing shell will cross previously collapsed shells just before reaching the pericenter\footnote{Here, we recall that due to the presence of the angular momentum, the shells do not reach the origin, but an orbital pericenter.}, and this will cause the enclosed mass to drop below $M$. We employ the scaled quantities $r'=r/R$, $m=M/M_0$, and $s=\Delta t/t_0$ and set the time origin at $t_0$ for the sake of simplicity. In this case, Eq.~\eqref{r_over_R} takes the following form
\begin{eqnarray}\label{r_p_eq}
     r'^{3/2}F(r',m)=\sqrt{2}(GM_{\rm J})^{3/2}m^{1/2}\left(0.15m^{2/3}-s\right),
\end{eqnarray}
where 
\begin{eqnarray}
    F(r',m)=\sqrt{2GM_{\rm J}-\dfrac{J_{0}^{2}}{R}\left(\dfrac{M_{0}}{M_{\rm J}}\right)^{1/3}\dfrac{m^{1/3}}{r'}}\times \hspace{0.5cm}\nonumber \\
    \left[GM_{\rm J}+\dfrac{J_{0}^{2}}{R}\left(\dfrac{M_{0}}{M_{\rm J}}\right)^{1/3}\dfrac{m^{1/3}}{r'}\right].
\end{eqnarray}
Also, correspondingly $0<s<0.15m^{2/3}$ and $m<1$. Specifically, in limit $r'\rightarrow [J_{0}^{2}(M_{0}/M_{\rm J})^{1/3}]/(2M_{\rm J}RG)$, it can be found that $m\rightarrow m_{\rm c}(s)=(s/0.15)^{3/2}$. This is the inverse of Eq.~\eqref{collapse_time}, which describes the collapse time of the initial mass shell $m$ as $s_{\rm c}(m)=0.15m^{2/3}\,\,$ \footnote{As can be seen in the following, for the parameters characteristic of the galaxy DD046, we have that $[J_{0}^{2}(M_{0}/M_{\rm J})^{1/3}]/(2M_{\rm J}RG)\simeq 0.001$. The fact that $r'$ cannot reach $0$ is due to the presence of the angular momentum.}.

Let's assume a mass shell that initially contains mass $M'$ and is in a critical state of collapse at the moment of $t=t_{0}+0.15(M'/M_{0})^{2/3}$. The equation of motion of this shell as it re-expands is
\begin{equation}
    \dfrac{d^{2}r}{dt^{2}}=-\dfrac{GM(r,t)}{r^{2}}+\dfrac{J^2}{r^3},
\end{equation}
which in scaled variables defined above equates to
\begin{equation} \label{r_prime_eq}
    \dfrac{d^{2}r'}{ds^{2}}=-\dfrac{2}{9}\dfrac{m(r',s)}{r'^{2}}+\dfrac{2}{9}\dfrac{J^{2}}{GM_{0}r'^{3}R}.
\end{equation}
Since the radius of the shell fulfills Eq.~\eqref{solution1} during the last phase of initial collapse, it can be assumed that the post-collapse motion acts as the time-reverse of the initial collapse. Therefore, the shell radius in scaled variables can be determined as follows
\begin{equation}\label{r_p_m_p_eq}
    r'^{3/2}F(r',m')=\sqrt{2}(GM_{\rm J})^{3/2}m'^{1/2}\left(s-s_{\rm c}(m')\right),
\end{equation}
where $m'=M'/M_{0}$, $s_{\rm c}(m')=0.15m'^{2/3}$, and
\begin{eqnarray}
    F(r',m')=\sqrt{2GM_{\rm J}-\dfrac{J_{0}^{2}}{R}\left(\dfrac{M_{0}}{M_{\rm J}}\right)^{1/3}\dfrac{m'^{1/3}}{r'}}\times \hspace{0.5cm}\nonumber \\
    \left[GM_{\rm J}+\dfrac{J_{0}^{2}}{R}\left(\dfrac{M_{0}}{M_{\rm J}}\right)^{1/3}\dfrac{m'^{1/3}}{r'}\right].
\end{eqnarray}
This can be used as initial conditions to solve Eq.~\eqref{r_prime_eq}. It can also be deduced from Eq.~\eqref{r_p_eq} that for all $r'$ and $s>s_{\rm c}(m')$, one obtains $m>m'$. This means that the deceleration of the shell during its re-expansion is larger than the acceleration it experiences during its first collapse to the center. Therefore, favorable conditions can be provided for the second collapse of the shell to the center at smaller radii. In fact, the final cusp in the halo density profile is the product of this asymmetry between the re-expansion and collapse of the mass shells. By redefining the variables as $s''=s/m'^{2/3}$, and $r''=r/m'^{7/9}$ and defining $f=m(r',s)/m'$, Eq.~\eqref{r_p_eq} can be determined as follows
\begin{eqnarray} \label{mio1}
    r''^{3/2}F(r'',m')=\sqrt{2}(GM_{\rm J})^{3/2}f^{1/2}\left(0.15f^{2/3}-s''\right),
\end{eqnarray}
in which
\begin{eqnarray}
F(r'',m')=\sqrt{2GM_J-\frac{J_0^2}{R} \left(\dfrac{M_{0}}{M_J}\right)^{1/3}\dfrac{1}{r'' m'^{4/9}}} \times \hspace{0.5cm}\nonumber \\
\left[ GM_J+\dfrac{J_0^2 }{R}\left(\dfrac{M_{0}}{M_J}\right)^{1/3}\dfrac{1}{r'' m'^{4/9}}\right].
\end{eqnarray}
Accordingly, Eq.\,\eqref{r_prime_eq} is specified as follows in terms of the variables scaled above
\begin{equation} \label{mio2}
    \dfrac{d^{2}r''}{ds''^{2}}=-\dfrac{2}{9}\dfrac{f(r'',s'')}{r''^{2}}+\dfrac{2}{9}\dfrac{J_{0}^{2}}{GM_{0}R}\dfrac{1}{r''^{3}m'^{4/9}}.
\end{equation}
Also,  at small radii, the initial solution of Eq.~\eqref{r_p_m_p_eq} is
\begin{equation} \label{mio3}
    r''^{3/2}F(r'',m')=\sqrt{2}(GM_{\rm J})^{3/2}\left(s''-0.15\right).
\end{equation}

Differently from \cite{2022MNRAS.517L..46W}, the post-collapse trajectories are dependent on $m'$ so they are not self-similar. In any case, the non-self-similarity is not strong, and it makes sense to integrate a shell's trajectory independently because they are not all coupled.

However, as we will show, the dependency of the density $\rho(r)$ from the radius is not that obtained by White, namely $\rho \propto r^{-12/7}$, since $r''$ is not a constant equal to $0.087$, but slightly depends on mass. 

By integrating numerically Eq. (\ref{mio2}), using the initial solution at a small radius, i.e., Eq. (\ref{mio3}), the function $f(r'',s'')$, i.e., Eq. (\ref{mio1}), and fixing the values of $R$, $M_0=M_J$, $J_0$, one can obtain the solution. In the case of $J_0=0$, the solution is the same as \cite{2022MNRAS.517L..46W}. In that case, the time and radius of the second apocentre are given by $s''=0.199$, and $r''=0.087$, which can be written in terms of the original variables $r$ and $t$ as in Eq.\,(14) of \cite{2022MNRAS.517L..46W}, i.e.,
\begin{equation} \label{eq:noJ}
r_{\rm max}=0.087 R \left(\dfrac{M'}{M_0}\right)^{7/9}.
\end{equation} 

In the case of nonzero angular momentum, one can obtain $r''=\left(0.104/m'^{0.1}\right)$. In other words, in the original variables, the radius of the apocentre can be written as 
\begin{equation} \label{eq:siiJ}
r_{\rm max}=\dfrac{0.104}{\left(M'/M_0\right)^{0.1}} R \left(\dfrac{M'}{M_0}\right)^{7/9}=0.104 R \left(\dfrac{M'}{M_0}\right)^{0.678}.
\end{equation}

In Fig. (\ref{fig:example_figure}), we have shown the post-collapse trajectories in the case of zero angular momentum (left panel), and in the presence of nonzero angular momentum and $m'=0.5$ (right panel). As can be seen from the figure, also from Eqs. (\ref{eq:noJ}), and (\ref{eq:siiJ}), $r_{\rm max}$ increases with $M'$, but slightly slower when the nonzero angular momentum is taken into account. The post-collapse equilibrium of the structure is reached in times much shorter than $t_0$, and it is established from the inside to the outside. 

In the presence of nonzero angular momentum, the mass in a radius $r$ scales as $M(r) \propto r^{1.48}$
%valore  0.0003
This dependence comes directly from Eq. \eqref{eq:siiJ}, solving with respect to $M(r)$.
This scaling can be obtained because the gravitational force at $r<r_{\rm max}$ has a small evolution at $t>t_{\rm max}$. Then $\rho(r)$ can simply be specified by calculating the ratio between the mass and the volume, leading to $\rho \propto \frac{M}{r^3} \propto \frac{r^{1.48}}{r^3} \propto r^{-1.52}$.

Interestingly, this result is in agreement with the $N$-body simulations, see, e.g. \citep[][]{2010ApJ...723L.195I, 2013JCAP...04..009A, 2014ApJ...788...27I, 2017MNRAS.471.4687A, 2018MNRAS.473.4339O, 2018PhRvD..97d1303D, 00001254, 2023MNRAS.518.3509D}.

We have to stress that the result $\rho^{-1.52}$, in agreement with \citep{2023MNRAS.518.3509D} simulations is obtained for, low, peculiar values of the specific angular momentum, namely $J_0$ (related to the product of velocity and radius), $R$, and $M$ has been fixed to those of DD046 \citep{Oh2015} (see their Fig.\,16, and Table 2). In the case of structures having a large specific angular momentum, as spiral galaxies similar to the Milky Way, and then a term $2 J_0^2/9 G M_0 R$ in Eq. (\ref{mio2}) larger than in the case of DD046, there will be a further flattening of the profile. About this issue, and the comparison with numerical simulations, see our Section III.

%
%As shown in \cite{2001ApJ...555..240B}, the specific angular momentum of each mass element is defined as $J=L/M=kr^{\alpha}$, %where $\alpha=1.1 \pm 0.3$ is a power-law index corresponding to the Gaussian distribution on dark matter halos, and $k=J_0/%R^{1.1 \pm 0.3}$, being $J_0$, and $R$, the typical specific angular momentum, and scale of a halo.

 Apart from the mainly used definition of angular momentum that we discussed, as shown in \cite{2001ApJ...555..240B}, the specific angular momentum of each mass element can be defined as $J=L/M=kr^{\alpha}$, where $\alpha=1.1 \pm 0.3$ is a power-law index corresponding to the Gaussian distribution on dark matter halos, and $k=J_0/R^{1.1 \pm 0.3}$, where $J_0$ and $R$ are typical specific angular momentum and scale of a halo, respectively.

In order to have an algebraically compact solution, one can choose $\alpha=0.9$. By repeating the calculations for the mentioned expression of angular momentum, one can recover the equations needed to obtain the post-collapse trajectories. Accordingly, the implicit equation for $f(r'',s'')$ is given by
\begin{eqnarray} \label{mio11}
    r''^{3/2}{}_{2}F_{1}\left(\dfrac{1}{2},\dfrac{15}{8};\dfrac{23}{8};\dfrac{5k^{2}(R r''m'^{7/9})^{0.8}}{Gm'M_{0}}\right)\hspace{1.5cm}\nonumber \\
    =f^{1/2}\left(0.15f^{2/3}-s''\right),
\end{eqnarray}
where ${}_{2}F_{1}(a,b;c;z)$ represents the hypergeometric function. The equation of motion, i.e., Eq. (\ref{mio2}), is specified as follows in terms of scaled variables
\begin{equation} \label{mio22}
    \dfrac{d^{2}r''}{ds''^{2}}=-\dfrac{2}{9}\dfrac{f(r'',s'')}{r''^{2}}+\dfrac{2}{9}\dfrac{k^{2}R^{0.8}}{GM_{0}}\dfrac{1}{r''^{1.2}m'^{0.377}}.
\end{equation}
Hence, the initial solution at a small radius takes the following form 
\begin{eqnarray} \label{mio33}
    r''\{{}_{2}F_{1}\left(\dfrac{1}{2},\dfrac{15}{8};\dfrac{23}{8};\dfrac{5k^{2}(R r''m'^{7/9})^{0.8}}{Gm'M_{0}}\right)\}^{2/3}\hspace{0.8cm}\nonumber\\
    =(s''-0.15)^{2/3}.
\end{eqnarray}
Similar to the method employed earlier and using Eqs. (\ref{mio11}), (\ref{mio22}), and (\ref{mio33}), the following formula can be obtained
\begin{equation} \label{eq:siiJ1}
r_{\rm max}=\dfrac{0.1138}{\left(M'/M_0\right)^{0.12}} R \left(\dfrac{M'}{M_0}\right)^{7/9}=0.1138 R \left(\dfrac{M'}{M_0}\right)^{0.658}.
\end{equation}

As a result, $\rho(r)$ can be specified simply by calculating the ratio between the mass and the volume, leading to  $\rho(r) \propto r^{-1.48}$. In \cite{2001ApJ...555..240B}, the specific angular momentum has a certain scatter, $J \propto r^{-1.1 \pm 0.3}$. Taking account of this scatter, the density profile is proportional to $\rho(r) \propto r^{(-1.44, -1.58)}$. 

Up to here, we have provided an analytical approach to determine the effect of angular momentum on the prompt cusp formed through gravitational collapse. In the next section, we will discuss the reasons why inclusion of the nonzero angular momentum produces a flatter profile than that obtained in \cite{2022MNRAS.517L..46W}. 

\section{Discussion} \label{sec:iii}

As noticed in \cite{2022MNRAS.517L..46W}, there are several points to discuss relative to the validity of the result obtained in that paper. The system used in that paper is spherically symmetric, then the motions are purely radial. In this condition, the density profile should have an inner slope close to or smaller than $-2$ \citep{1972ApJ...176....1G,1984ApJ...281....1F,Bertschinger1985,Hoffman1985}. The scaling argument in \cite{2022MNRAS.517L..46W} fails unless some angular momentum is acquired before particles reach the final orbits. As previously discussed, angular momentum can be acquired through tidal torques experienced by protohalos \citep{Peebles1969,White1984,Catelan1996}, 
%or by random velocities (non-radial motions) \cite{Ryden1987}, 
or deviation from spherical symmetry. According to \cite{2022MNRAS.517L..46W}, angular momentum would restore the slope $-12/7$ for the cusp (as expected from a model proposed by \cite{White1996}), but in the case, it is too strong it could invalidate the result. Again, as previously reported, several studies \citep{Ryden1987,AvilaReese1998,Nusser2001,Hiotelis2002,LeDelliou2003,Ascasibar2004,Williams2004,2009ApJ...698.2093D,Polisensky2015} arrived to the conclusion that large amounts of angular momentum, 
%ordered or random, 
leads to shallower inner density profiles, till even the formation of a central core. As mentioned by \cite{2022MNRAS.517L..46W}, the main point leading to doubt of the applicability of the argument used in \cite{2022MNRAS.517L..46W} is the assumption of spherical symmetry. However, the same author after an argumentation of this issue arrives to conclude that $\rho \propto r^{-12/7}$, even if as shown in \citep{2023MNRAS.518.3509D} the initial collapse is complex and very far from spherical. However, simulations find a slope $\simeq -1.5$, flatter than that obtained in \cite{2022MNRAS.517L..46W}. The author then asks whether the model captures the features of violent relaxation in the inner region of the peak, or whether there are some factors that explain the difference between the simulated slopes of $-1.5$ and $-12/7$ obtained in that paper.
As shown in \cite{2009ApJ...698.2093D}, there are several factors that change the inner slope of the density profile. In that paper, the collapse was studied taking into account ordered and random angular momentum, dynamical friction \citep{DelPopolo1996}, dark energy, and dark matter contraction due to baryonic infall \citep{Blumenthal1986,Gnedin2004}. Those physical effects influence the structure formation and the inner slope in different ways. For example, baryonic infall produces a steepening of the profile, while angular momentum and dynamical friction slow down the collapse, and flatten the profile. In the present paper, we have decided to take into account only the ordered angular momentum to show how it is enough to reduce the inner slope of the density profile in agreement with theoretical studies and simulations
\citep{Ryden1987,AvilaReese1998,Nusser2001,Hiotelis2002,LeDelliou2003,Ascasibar2004,Williams2004,2009ApJ...698.2093D,Polisensky2015}. In our model, the change of the inner structure is related to the interaction of the structure studied with the neighboring ones, arising from the asphericity of those structures (see \cite{DelPopolo1999} for a discussion on the relation between angular momentum acquisition, asphericity, and structure formation). Asphericity gives rise to a mass-dependent inner slope. The equation of motion in our model contains a mass-dependent angular momentum, born from the quadrupole momentum of the proto-structures with the tidal field of the neighboring objects. This term slightly breaks the self-similarity of the trajectories of the mass shells. 
Hence, the turnaround epoch and collapse time would change. The collapse in our model is different from that of \cite{2022MNRAS.517L..46W}. Both turnaround epoch and collapse time change, together with the collapse threshold $\delta_c$, which becomes mass-dependent and a monotonic decreasing function of the mass (see Fig. 1 in \cite{DelPopolo2017})\footnote{We want also to recall that the behavior of the threshold implies that less massive perturbation (e.g. galaxies) to form structures must cross a higher threshold than more massive ones. Using the peak formalism, considering the peak height $\nu=\delta_{\rm c}/\sigma(M)$ (being $\sigma$ the mass variance), the angular momentum acquired by a peak is proportional to the turnaround time, $t_{\rm ta}$, and anti-correlated with the peak height, $J \propto t_{\rm ta} \propto \nu^{-3/2}$ \cite{2009ApJ...698.2093D,Hoffman1986,Polisensky2015}. Since low peaks acquire a larger angular momentum with respect to high peaks, they must have a higher density contrast in order to collapse and form structure \citep{2009ApJ...698.2093D,DelPopolo1998,DelPopolo2001,DelPopolo2002,Peebles1990}}.  

The flattening of the profile can be explained as follows. In the case of pure radial orbits, the inner part of the profile is dominated by particles from the outer shells. When the angular momentum increases, these particles are closer to the maximum radius, and this gives rise to a shallower profile. Particles having smaller angular momentum will enter the inner part (core) of the halo, but with a reduced radial velocity in comparison with purely radial collapse. Some particles have an angular momentum so large that they never fall in the core. In other terms, particles with larger angular momentum are prevented from coming close to the central region of the halo, then contributing to the central density. Consequently, the profile is flattened. Moreover, this result is in agreement with the previrialization conjecture \cite{Peebles1990}, according to which initial asphericities and tidal interactions between neighboring protostructures give rise to non-radial motions opposing the collapse. Apart from this, the role of angular momentum in flattening the profile is in agreement with previously mentioned studies. 

One of the main points mentioned in \cite{2022MNRAS.517L..46W} is that the difference between the prediction of the analytical approach and the simulations may be due to the effect of some additional factors. To address this point, as shown in \cite{2009ApJ...698.2093D}, it should be noted that the effect of additional factors on the distribution of the inner regions of halos is a potential possibility. In this work, we have shown that the consideration of angular momentum affects the slope of the density profile, in such a way that the difference between the prediction obtained from the theoretical approach and the simulations is significantly reduced. 

Before concluding, as we previously wrote, we recall that the result $\rho^{-1.52}$, in agreement with simulations, is obtained for, low, peculiar values of the specific angular momentum, radius, and mass. A further flattening with respect to \cite{2022MNRAS.517L..46W} should be waited for large spiral galaxies like the Milky Way.  
As we reported in the introduction, the high-resolution simulation by \citep{2023MNRAS.518.3509D}
have dark matter only, so baryon-induced effects like, for example, adiabatic contraction \citep{Blumenthal1986,Gnedin2004} do not apply. Limiting ourselves to this issue, the effect of adiabatic contraction is that of steepening the profile. In other terms, the slope $\gamma \approx -1.5$ in \citep{2023MNRAS.518.3509D} could be modified by the effects not taken into account. Also, our model is not taking into account the adiabatic contraction. Then, to get a more precise value of the slope, it will be important to run appropriate simulations, and in our case to use a model like that described in \citep{2009ApJ...698.2093D}
taking not only into account angular momentum but also dynamical friction, adiabatic contraction, etc.

%%%%%%%%%%%%%%%%%%%%%%%%%%%%%%%%%%%%%%%%%%%%%%%%%%
\section{Conclusions} \label{sec:iv}
In this paper, we have extended the model proposed by \cite{2022MNRAS.517L..46W}, relative to the post-collapse evolution of density peaks, looking at the effect angular momentum can have on the author's final solution. In particular, we wanted to see if angular momentum could reduce the discrepancy between the density profile extracted from \cite{2022MNRAS.517L..46W} and that obtained from simulations. As several times cited, several papers stressed that angular momentum has the effect of flattening the inner slope of density profiles. By modifying the equations presented in \cite{2022MNRAS.517L..46W}, and including the nonzero angular momentum, we have shown that on a timescale smaller than the peak collapse, $t_0$, the equilibrium configuration of the peak is a cusp but with a flatter slope $\rho \propto r^{-1.52}$, for the classical form of the specific angular momentum, $J \propto M^{2/3}$. The previous result indicates how angular momentum can reduce the discrepancy between the slope of the density profile derived in \cite{2022MNRAS.517L..46W} and that obtained in   
in high-resolution numerical simulations, namely $\rho \propto r^{\alpha}$ ($\alpha \simeq -1.5$). 
The reason why the angular momentum flattens the inner density profile is qualitatively justified by the fact that in the case we have considered a collapse with pure radial orbits, as in \cite{2022MNRAS.517L..46W}, outer particles dominate the inner part of the profile, and this gives rise to cuspier density profiles. If the nonzero angular momentum is present, the particles' orbits are closer to the maximum radius, with the consequence that a flatter profile can be obtained. 
In other terms, particles with larger angular momentum are prevented from coming close to the halo's center, then contributing to the central density. Consequently, the density profile is flattened.

\section*{ACKNOWLEDGMENTS}
The authors would like to gratefully acknowledge Prof. Giovanni Russo from the mathematical department of Catania University for helping to advance some of the calculations.

\bigskip

%\newpage

\bibliography{num-rel2}

\end{document}